\documentclass[aps,pra,reprint,superscriptaddress,showpacs]{revtex4-1}

\usepackage{latexsym}
\usepackage{graphics}
\usepackage{amsmath}
\usepackage{color}
\usepackage{graphicx}

\newcommand {\bfp} {{\bf p}}
\newcommand {\bfq} {{\bf q}}

\renewcommand {\d} {{\rm d}}

\newcommand {\E} {\varepsilon}

\newcommand {\om} {\omega}
\newcommand {\Om} {\Omega}

\begin{document}

\title{Comparative analysis of the secondary electron yield from carbon nanoparticles
and pure water medium}

\author{Alexey Verkhovtsev}
\email[]{verkhovtsev@mail.ioffe.ru}
\affiliation{MBN Research Center, Altenh\"oferallee 3, 60438 Frankfurt am Main, Germany}
\affiliation{A.F. Ioffe Physical-Technical Institute, Politekhnicheskaya ul. 26,
194021 St. Petersburg, Russia}

\author{Sally McKinnon}
\affiliation{Centre for Medical Radiation Physics (CMRP),
University of Wollongong, NSW, Australia}

\author{Pablo de Vera}
\affiliation{Departament de F\'{\i}sica Aplicada, Universitat d'Alacant, Alicante, Spain}
\affiliation{Department of Physical Sciences, The Open University, Milton Keynes, UK}

\author{Eugene Surdutovich}
\affiliation{Department of Physics, Oakland University, Rochester, Michigan 48309, USA}

\author{Susanna Guatelli}
\affiliation{Centre for Medical Radiation Physics (CMRP),
University of Wollongong, NSW, Australia}
\affiliation{Illawarra Health and Medical Research Institute (IHMRI),
University of Wollongong, NSW, Australia}

\author{Andrei V. Korol}
\affiliation{MBN Research Center, Altenh\"oferallee 3, 60438 Frankfurt am Main, Germany}
\affiliation{Department of Physics, St. Petersburg State Maritime Technical University,
Leninskii pr. 101, 198262 St. Petersburg, Russia}

\author{Anatoly Rosenfeld}
\affiliation{Centre for Medical Radiation Physics (CMRP),
University of Wollongong, NSW, Australia}
\affiliation{Illawarra Health and Medical Research Institute (IHMRI),
University of Wollongong, NSW, Australia}

\author{Andrey V. Solov'yov}
\affiliation{MBN Research Center, Altenh\"oferallee 3, 60438 Frankfurt am Main, Germany}
\affiliation{A.F. Ioffe Physical-Technical Institute, Politekhnicheskaya ul. 26,
194021 St. Petersburg, Russia}


\begin{abstract}
The production of secondary electrons generated by carbon nanoparticles and
pure water medium irradiated by fast protons is studied by means of model
approaches and Monte Carlo simulations.
It is demonstrated that due to a prominent collective response to an external
field, the nanoparticles embedded in the medium enhance the yield of low-energy
electrons.
The maximal enhancement is observed for electrons in the energy range where
plasmons, which are excited in the nanoparticles, play the dominant role.
Electron yield from a solid carbon nanoparticle composed of fullerite,
a crystalline form of C$_{60}$ fullerene, is demonstrated to be several
times higher than that from liquid water.
Decay of plasmon excitations in carbon-based nanosystems thus represents a
mechanism of increase of the low-energy electron yield, similar to the case
of sensitizing metal nanoparticles.
This observation gives a hint for investigation of novel types of sensitizers
to be composed of metallic and organic parts.
\end{abstract}

\maketitle

\section{Introduction}

Radiotherapy is currently one of the most frequently used technologies to treat
tumors, which are a major health concern \cite{Delaney_2005_Cancer.104.1129}.
However, this technique has a limitation which comes from the sensitivity of
healthy tissues, surrounding the tumor, to radiation.
To make the treatment more efficient, one needs to minimize the dose delivered
to the healthy tissue, thus preventing harmful effects of radiation exposure.
Therefore, approaches that enhance radiosensitivity within tumors relative to
normal tissues have the potential to
become advantageous radiotherapies.
A search for such approaches is within the scope of several ongoing
multidisciplinary projects \cite{COST_Nano-IBCT,ARGENT}.

One of the most promising modern treatment techniques is ion-beam cancer therapy
(IBCT) \cite{Bacarelli_2010_EPJD.60.1, Schardt_2010_RevModPhys.82.383,
Durante_2010_NatRevClinOncol.7.37}.
In this technique, radiation damage is initiated by fast ions incident on tissue.
Propagating through the medium, the projectiles deposit their
kinetic energy due to the ionization and excitation processes.
Biodamage due to ionizing radiation involves a number of phenomena,
which happen on various spatial, time, and energy scales.
The key phenomena can be described within the so-called multiscale
approach to the physics of radiation damage with ions
(see reference \cite{Surdutovich_2014_EPJD_Colloquia_Paper} and references therein).
As a result of the interaction of projectiles with the medium,
secondary particles, such as electrons, free radicals, etc.,
are produced.
By now, it is generally accepted that the vast portion of biodamage
done by incident heavy ions is related to these secondary particles
\cite{Surdutovich_2014_EPJD_Colloquia_Paper, Solov'yov_2009_PhysRevE.79.011909,
Michael_2000_Science.287.1603, Denifl_RadDamage_Chapter}.
Particularly, the low-energy electrons, having the kinetic energy from
a few eV to several tens of eV, have been shown to act as important
agents of biodamage
\cite{Boudaiffa_2000_Science.287.1658, Huels_2003_JAmChemSoc.125.4467}.

Metallic nanoparticles, especially those composed of noble metals,
were proposed recently to act as sensitizers in cancer treatments
with ionizing radiation
\cite{Herold_2000_IntJRadiatBiol.76.1357, Hainfeld_2004_PhysMedBiol.49.N309,
McMahon_2011_SciRep.1.18, Chen_2006_JNanosciNanotechnol.6.1159,
Porcel_2010_Nanotechnology.21.085103}.
Such nanoagents delivered to the tumor region can boost the production
of secondary electrons near the target
\cite{Sanche_2011_Nanotechnology.22.465101, Sanche_2008_RadiatRes.169.19}.
The enhanced production of low-energy electrons will also lead to an
increase in the number of free radicals \cite{Sicard-Roselli_2014_Small.10.3338}
as well as other reactive species, like hydrogen peroxide ${\rm H}_2{\rm O}_2$
\cite{Porcel_2014_NanomedNanotechBiolMed}, which can propagate from the
cytoplasm to the cell nucleus.
Thus, these species can deliver damaging impacts onto the DNA from the radiation
induced damages associated with the presence of nanoparticles in other cell
compartments \cite{Stefancikova_2014_CancerNanotech.5.6}.
An enhanced production of the secondary species will lead to an increase of
the relative biological effectiveness of ionizing radiation.
This quantity is defined as the ratio of the dose delivered by photons
to that delivered by different radiation modalities, leading to the
same biological effects, such as the probability of an irradiated cell
inactivation.

The physical mechanisms of enhancement of the electron yield from
sensitizing nanoparticles are still a debated issue.
In the recent studies \cite{Verkhovtsev_2014_NPs, Verkhovtsev_2014_GNP},
it was discovered that a significant increase in the number of emitted
electrons due to irradiation of noble metal nanoparticles by fast ions
comes from the two distinct types of collective electron excitations.
It was demonstrated that the yield of the $1 - 10$~eV electrons is
strongly enhanced due to the decay of plasmons, i.e. collective excitations
of delocalized valence electrons in metallic nanoparticles.
For electron energies of about $10 - 30$~eV, the dominating contribution
to the electron yield arises from the atomic giant resonance associated
with the excitation of $d$-electrons in individual atoms in a nanoparticle
\cite{Verkhovtsev_2014_NPs}.

Excitation of plasmons by time-dependent external electric fields is a
characteristic feature of not only metallic but also, to some extent,
of carbon nanoscale systems.
For instance, it is generally accepted that plasmon excitations dominate
the spectra of photo- and electron impact ionization of fullerenes
\cite{Hertel_1992_PhysRevLett.68.784, Lambin_Lukas_1992_PhysRevB.46.1794,
Mikoushkin_1998_PhysRevLett.81.2707, Verkhovtsev_2012_JPhysB.45.141002,
Kafle_2008_JPhysSocJpn.77.014302, Verkhovtsev_2013_PhysRevA.88.043201}
and polycyclic aromatic hydrocarbons (PAHs)
\cite{Ling_1996_ChemPhysLett.257.587, Verkhovtsev_2014_JPCS.490.012159}.

In this paper, we demonstrate that the decay of plasmons excited in carbon
nanoparticles also plays a prominent role in the production of low-energy electrons.
Due to the collective response to a time-dependent external electric field,
these systems enhance the production of secondary electrons in a biological medium,
in the energy range where the plasmons play the dominant role.
This is done by the calculation of spectra of secondary electrons
ejected from a carbon nanoparticle composed of fullerite, a crystalline
form of C$_{60}$ fullerene, irradiated by fast protons.
The contribution of plasmon excitations to the electron production
is evaluated by means of the plasmon resonance approximation
\cite{Kreibig_Vollmer, Connerade_AS_PhysRevA.66.013207,
Solovyov_review_2005_IntJModPhys.19.4143, Verkhovtsev_2012_EPJD.66.253}.
The results of these calculations are compared to the model calculations
based on the dielectric formalism \cite{Lindhard_1954_dielectric_formalism}
and Monte Carlo simulations
\cite{Agostinelli_2003_NIMA.506.250, Allison_2006_IEEETNS.53.270}, carried out
for pure water medium and for the medium with an embedded carbon nanoparticle.
Utilizing and comparing different theoretical and numerical methods,
we provide a recipe for evaluation of the electron production in the
kinetic energy range from a few eV to thousands of eV.
A single method does not allow one to properly quantify the secondary
electron yield in a broad energy range; thus, a combination of different
approaches is required.

\section{Theory and Computational Details}

\subsection{Plasmon resonance approximation}
\label{PRA_theory}

The contribution of collective electron excitations to the ionization
spectra of carbon nanoparticles is evaluated by means of the
plasmon resonance approximation (PRA)
(see references \cite{Kreibig_Vollmer, Connerade_AS_PhysRevA.66.013207,
Solovyov_review_2005_IntJModPhys.19.4143, Verkhovtsev_2012_EPJD.66.253}
and references therein).
This approach postulates that the dominating contribution to the
ionization cross section in the vicinity of the plasmon resonance
comes from collective electron excitations, while single-particle
effects give a small contribution compared to the collective modes
\cite{Gerchikov_1997_JPhysB.30.5939, Gerchikov_2000_PhysRevA.62.043201}.
In the past, this approach has provided a clear physical explanation
of the resonant-like structures in photoionization spectra
\cite{Verkhovtsev_2013_PhysRevA.88.043201, Connerade_AS_PhysRevA.66.013207}
and differential inelastic scattering cross sections
\cite{Mikoushkin_1998_PhysRevLett.81.2707, Verkhovtsev_2012_JPhysB.45.141002,
Gerchikov_1997_JPhysB.30.4133, Bolognesi_2012_EurPhysJD.66.254}
of metallic clusters and carbon fullerenes irradiated by photons and
fast electrons.

To start with, we evaluate the plasmon contribution to the ionization
spectrum of an isolated C$_{60}$ molecule.
Within the utilized model, the fullerene is represented as a spherical
''jellium'' shell of a finite width, $\Delta R = R_2 - R_1$,
so the electron density is homogeneously distributed over the shell with
thickness $\Delta R$
\cite{Lambin_Lukas_1992_PhysRevB.46.1794, Oestling_1993_EurophysLett.21.539,
Lo_2007_JPhysB.40.3973}.
The chosen value, $\Delta R = 1.5~{\rm \AA}$, corresponds to the size of
the carbon atom \cite{Oestling_1993_EurophysLett.21.539}.

The interaction of a hollow system with a non-uniform electric field, created
in collisions with charged projectiles, leads to the time-dependent variation
of the electron density appearing on the inner and outer surfaces of the hull
as well as in its interior \cite{Verkhovtsev_2012_EPJD.66.253}.
This variation leads to the formation of a surface plasmon, which has two normal
modes, the symmetric and antisymmetric
\cite{Lambin_Lukas_1992_PhysRevB.46.1794, Oestling_1993_EurophysLett.21.539,
Lo_2007_JPhysB.40.3973, Korol_AS_2007_PhysRevLett_Comment},
and of a volume plasmon \cite{Gerchikov_2000_PhysRevA.62.043201}, which occurs
due to a local compression of the electron density inside the shell.
The detailed explanation of formation of different plasmon modes can be found
in references~\cite{Connerade_AS_PhysRevA.66.013207, Verkhovtsev_2012_EPJD.66.253}.


The utilized approach relies on several parameters, which include the oscillator
strength of the plasmon excitation, position of the plasmon resonance peak, and
its width.
The choice of these parameters can be justified by comparing the model-based
spectra with either experimental data or the results of more advanced
{\it ab initio} calculations.
As a benchmark of the utilized approach, the photo- and electron impact
ionization cross sections of carbon-based systems, namely fullerenes and PAHs,
were calculated recently
\cite{Verkhovtsev_2012_JPhysB.45.141002, Verkhovtsev_2013_PhysRevA.88.043201,
Verkhovtsev_2014_JPCS.490.012159, Bolognesi_2012_EurPhysJD.66.254}.
The results obtained for C$_{60}$
\cite{Verkhovtsev_2012_JPhysB.45.141002, Verkhovtsev_2013_PhysRevA.88.043201,
Bolognesi_2012_EurPhysJD.66.254}
agreed well with experimental data on photoionization
\cite{Kafle_2008_JPhysSocJpn.77.014302} and electron inelastic scattering
\cite{Verkhovtsev_2012_JPhysB.45.141002, Bolognesi_2012_EurPhysJD.66.254}.
Being a clear physical model which describes collective electron excitations,
the PRA has been proven to be a useful tool for interpretation of experimental
results and making new numerical estimates.


Within the PRA, the double differential inelastic scattering cross section of
a fast projectile in collision with a hull-like system can be defined as a sum
of three terms \cite{Verkhovtsev_2012_JPhysB.45.141002, Verkhovtsev_2012_EPJD.66.253}
(hereafter, we use the atomic system of units, $m_e = |e| = \hbar = 1$):
\begin{equation}
\frac{\d^2\sigma_{\rm pl}}{\d\E_2 \d\Om_{{\bfp}_2}} =
\frac{\d^2\sigma^{(s)} }{\d\E_2 \d\Om_{{\bfp}_2}} +
\frac{\d^2\sigma^{(a)} }{\d\E_2 \d\Om_{{\bfp}_2}} +
\frac{\d^2\sigma^{(v)} }{\d\E_2 \d\Om_{{\bfp}_2}} \ ,
\label{Equation.01}
\end{equation}

\noindent
which describe the partial contribution of the surface (the two modes, $s$ and $a$)
and the volume ($v$) plasmons.
Here $\E_2$ is the kinetic energy of the scattered projectile, ${\bfp}_2$ its momentum,
and $\Om_{{\bfp}_2}$ its solid angle.
The cross section $\d^2\sigma_{\rm pl} / \d\E_2 \d\Om_{{\bfp}_2}$ can be written
in terms of the energy loss $\Delta \varepsilon = \E_1 - \E_2$, of the incident
particle of energy $\E_1$.
Integration of $\d^2\sigma_{\rm pl} / \d\Delta\E \, \d\Om_{{\bfp}_2}$ over the
solid angle leads to the single differential cross section:
\begin{equation}
\frac{{\rm d}\sigma_{\rm pl}}{{\rm d}\Delta\E} =
\int {\rm d}\Omega_{{\bf p}_2}
\frac{{\rm d}^2\sigma_{\rm pl}}{{\rm d}\Delta \varepsilon \, {\rm d}\Omega_{{\bf p}_2}}
= \frac{2\pi}{p_1 p_2} \int\limits_{q_{\rm min}}^{q_{\rm max}} q \, {\rm d}q
\frac{{\rm d}^2\sigma_{\rm pl}}{{\rm d}\Delta \varepsilon \, {\rm d}\Omega_{{\bf p}_2}} \ ,
\end{equation}

\noindent
where
${\bfp}_1$ is the initial momentum of the projectile and
${\bfq} = {\bfp}_1 - {\bfp}_2$ is the transferred momentum.
Explicit expressions for the contributions of the surface and volume plasmons,
entering equation~(\ref{Equation.01}), obtained within the plane-wave Born
approximation, are presented in reference~\cite{Verkhovtsev_2012_EPJD.66.253}.
The Born approximation is applicable since the considered collision
velocities ($v_1 = 2 - 20$~a.u.) substantially exceed the characteristic
velocities of delocalized electrons in the fullerene ($v_e \approx 0.7$~a.u.).

The surface and volume plasmon terms appearing on the right-hand side of
equation~(\ref{Equation.01}) are constructed as a sum over different multipole
contributions corresponding to different values of the angular momentum $l$:
\begin{eqnarray}
\begin{array}{l l}
\displaystyle{ \frac{\d^2\sigma^{(i)} }{\d\E_2 \d\Om_{{\bfp}_2}} }
\propto
\sum\limits_{l} \frac{ \om_{l}^{(i)2}\, \Gamma_{l}^{(i)}  }
{ \bigl(\om^2-\om_{l}^{(i)2}\bigr)^2 + \om^2\Gamma_{l}^{(i)2} }
\vspace{0.2cm} \\
\displaystyle{ \frac{\d^2\sigma^{(v)} }{\d\E_2 \d\Om_{{\bfp}_2}} }
\propto
\sum\limits_{l} \frac{ \om_p^2\, \Gamma_l^{(v)} }
{ \bigl(\om^2-\om_p^2\bigr)^2+\om^2\Gamma_l^{(v)2} } \ ,
\end{array}
\label{Equation.02}
\end{eqnarray}

\noindent
where $i = s,a$ denotes the two modes of the surface plasmon.
Their frequencies are given by
\cite{Verkhovtsev_2012_EPJD.66.253, Oestling_1993_EurophysLett.21.539}:
\begin{equation}
\displaystyle{
\om_{l}^{(s/a)}
=
\left( 1 \mp  \frac{1}{2l+1} \sqrt{1 + 4l(l+1)\xi^{2l+1}} \right)^{1/2}
\frac{\om_p}{\sqrt{2}}
}
\label{MultipoleVariation.2a}
\end{equation}

\noindent
where '$-$' and '$+$' stand for symmetric $(s)$ and antisymmetric $(a)$ modes,
respectively, and $\xi = R_1/R_2$ is the ratio of the inner to the outer radii
of the shell.
The volume plasmon frequency $\om_p$, associated with the ground-state electron
density $\rho_0$, is given by
\begin{equation}
\om_p = \sqrt{4 \pi \rho_0} = \sqrt{ \frac{3N}{R_2^3 (1 - \xi^3)} } \ ,
\label{eq_02}
\end{equation}

\noindent
where $N$ is the number of delocalized electrons involved in the collective excitation.
In the case of a fullerene C$_n$, the number $N$ of delocalized electrons represents
the four 2$s^2$2$p^2$ valence electrons from each carbon atom.
Thus, we assume that 240 delocalized electrons of C$_{60}$ contribute to the formation
of plasmons.

In reference~\cite{Gerchikov_1997_JPhysB.30.4133} it was shown that the excitations
with large angular momenta have a single-particle rather than a collective nature.
With increasing $l$, the wavelength of plasmon excitation, $\lambda_{\rm pl} = 2\pi R/l$,
becomes smaller than the characteristic wavelength of the delocalized electrons
in the system, $\lambda_e = 2\pi /\sqrt{2\epsilon}$.
Here $\epsilon$ is the characteristic electron excitation energy in the cluster,
$\epsilon \sim I_p$, and $I_p$ is the ionization threshold of the system
($I_p({\rm C_{60}}) \sim 7.5$~eV \cite{Hertel_1992_PhysRevLett.68.784}).
In the case of the C$_{60}$ fullerene, the estimates show that the excitations with
$l > 3$ are formed by single electron transitions rather than by the collective ones.
Therefore, only terms corresponding to the dipole ($l = 1$), quadrupole ($l = 2$)
and octupole ($l = 3$) plasmon terms have been accounted for in the sum over $l$
in equation~(\ref{Equation.02}).
\begin{table}[h]
\centering
\caption{
Peak positions of the surface and the volume plasmon modes as well as
their widths used in the present calculations.
All values are given in eV.}
\begin{tabular}{p{1cm}p{1.25cm}p{1.25cm}p{1.25cm}p{1.25cm}}
\hline
            &  $l = 1$  &  $l = 2$  &  $l = 3$  \\
\hline
$\om_l^{(s)}$  &     19.0   &    25.5   &    30.5   \\
$\Gamma_{l}^{(s)}$  &     11.4   &    15.3   &    18.3   \\
$\om_l^{(a)}$  &    33.2   &    31.0   &    29.5   \\
$\Gamma_l^{(a)}$  &     33.2   &    31.0   &    29.5   \\
$\om_p$     & \multicolumn{3}{c}{37.1} \\
$\Gamma_l^{(v)}$  & \multicolumn{3}{c}{37.1} \\
\hline
\end{tabular}
\label{Table1}
\end{table}

Following the methodology utilized in reference~\cite{Bolognesi_2012_EurPhysJD.66.254},
we assume that the ratio $\gamma_l = \Gamma_l/\om_l$ of the width of the
plasmon resonance to its frequency equals to $\gamma_l^{(s)} = 0.6$ for
all multipole terms of the symmetric mode, and to $\gamma_l^{(a)} = 1.0$
for the antisymmetric mode.
These values were utilized previously to describe experimental data on
photoionization \cite{Kafle_2008_JPhysSocJpn.77.014302} and electron inelastic
scattering \cite{Verkhovtsev_2012_JPhysB.45.141002, Bolognesi_2012_EurPhysJD.66.254}
of gas-phase C$_{60}$.
The value $\gamma_l^{(s)} = 0.6$ is also close to the numbers obtained from the earlier
photoionization and electron energy loss experiments on neutral C$_{60}$
\cite{Hertel_1992_PhysRevLett.68.784, Mikoushkin_1998_PhysRevLett.81.2707}.
The value $\gamma_l^{(a)} = 1.0$ is consistent with the widths of the second plasmon
resonance observed in the photoionization of C$_{60}^{q+}$ ($q = 1 - 3$) ions
\cite{Scully_2005_PhysRevLett.94.065503}.
For the volume plasmon, we consider the ratio $\gamma_l^{(v)} = \Gamma_l^{(v)}/\om_p = 1.0$.
The values of the plasmon resonance peaks and the widths are summarized
in Table~\ref{Table1}.

\subsection{Dielectric formalism}

The secondary electron production in a pure water medium
as well as in a carbon nanoparticle
was investigated
by means of a model approach based on the dielectric formalism
\cite{Lindhard_1954_dielectric_formalism}.
This method relies on experimental measurements of the energy-loss
function of the target medium, ${\rm Im}[-1/\epsilon(\om, q)]$, where
$\epsilon(\om, q)$ is the complex dielectric function, with $\om$ and $q$
being the energy and the momentum transferred to the electronic excitation,
respectively.
In reference~\cite{Scifoni_2010_PhysRevE.81.021903}, this approach was used to
obtain spectra of secondary electrons generated in liquid water by energetic ions.
An alternative method to calculate the impact ionization cross sections of various
biological media was proposed recently
\cite{deVera_2013_PhysRevLett.110.148104, deVera_2014_PhysRevLett}.
Instead of calculating the exact energy-loss function and ionization threshold
for different electronic shells of a molecule composing the target medium,
this approach aims at calculating the mean value of the binding energies for several
outer shells.
It is assumed that ionization of these shells happens if the energy transferred
to the medium exceeds this mean value of the binding energies
\cite{deVera_2013_PhysRevLett.110.148104}.
The formalism presented allows one to calculate the cross sections not
only for liquid water but also for a real biological medium containing
sugars, amino acids, etc.
In particular, it was utilized recently \cite{deVera_2014_EPJD.68.96}
to study ionization and energy deposition in different subcellular compartments,
such as cell nucleus and cytoplasm, due to proton irradiation.
In this work, we apply this formalism to study the electron production from
a nanoparticle composed of fullerite.

\subsection{Monte Carlo simulation  of secondary electron yield}

Monte Carlo simulations of secondary electron production in a nanoparticle
were performed using Geant4, version 9.6 patch 1
\cite{Agostinelli_2003_NIMA.506.250, Allison_2006_IEEETNS.53.270}.
The simulation geometry consisted of a 50 nm diameter spherical nanoparticle
of variable material placed at the center of a 5 $\mu$m world of liquid water.
A 4 $\mu$m sided cube was included to allow the use of different secondary particle
production thresholds in different regions in order to optimize execution times.
Monoenergetic protons propagating from a point source were incident from the edge
of the nanoparticle.

The material of the nanoparticle was simulated as liquid water or a customized
fullerene material alternatively.
The fullerene material properties were set by scaling the density of the Geant4
element carbon according to the calculated density of a face-centered cubic (fcc)
structure of fullerite.

The Low Energy Electromagnetic Physics Package \cite{Physics_RefManual_2012},
using the Livermore Data Libraries, was selected to model the interactions of
electrons and photons in the nanoparticle.
Models describing proton interactions in the nanoparticle were selected following
the Geant4 advanced example ''Microdosimetry''.
The ionization model implemented for protons was the Geant4 ''BraggIonGas'' model,
valid for protons kinetic energy up to 2~MeV, while the Bethe-Bloch model was
adopted for higher energies.
In the nanoparticle, nuclear stopping power was modeled using the Geant4
''ICRU49NucStopping'' model.
The multiple scattering was modeled for all charged particles with the Geant4
''UrbanMsc95'' model \cite{Physics_RefManual_2012}.
Atomic de-excitation (fluorescence and Auger electrons) was modeled as well
\cite{Guatelli_2007_IEEETNS.54.585}.
Secondary electron production from the nanoparticle is limited to the electrons
with kinetic energy greater than 250 eV as this is the low-energy limit of
validity of the Livermore Data Libraries \cite{Chauvie_2004_ProcIEEE-NSS}.

The Geant4-DNA Very Low Energy extensions \cite{Incerti_2010_MedPhys.37.4692} were
adopted in liquid water surrounding the nanoparticle to model in detail particle
interactions down to a few eV scale.
Physical interactions modeled for protons in the water sphere were G4DNAExcitation,
G4DNAIonisation, and G4DNAChargeDecrease.
The models used are the default Geant4-DNA model classes.

The simulations in this study modeled the interactions of 1~MeV protons generated
from one position and in one direction incident on the nanoparticle.
Secondary electrons were produced in the nanoparticle with a cut of 250 eV.
The cut is the threshold of production of secondary particles.
Below the cut, secondary electrons are not produced and their energy is deposited
locally, while above the cut, secondary electrons are produced and tracked in the
nanoparticle and in the surrounding medium.
The kinetic energy spectra of secondary electrons escaping the nanoparticle were
retrieved and the spectra at creation was compared directly to the same physical
quantity calculated by means of the analytical model.
The proportion of escaping secondary electrons produced within the fullerite-like
nanoparticle was 98.5\%.

\section{Results and Discussion}
\subsection{Electron production by an isolated C$_{60}$ molecule due to
the plasmon excitation mechanism}

\begin{figure}[ht]
\centering
\includegraphics[width=0.45\textwidth,clip]{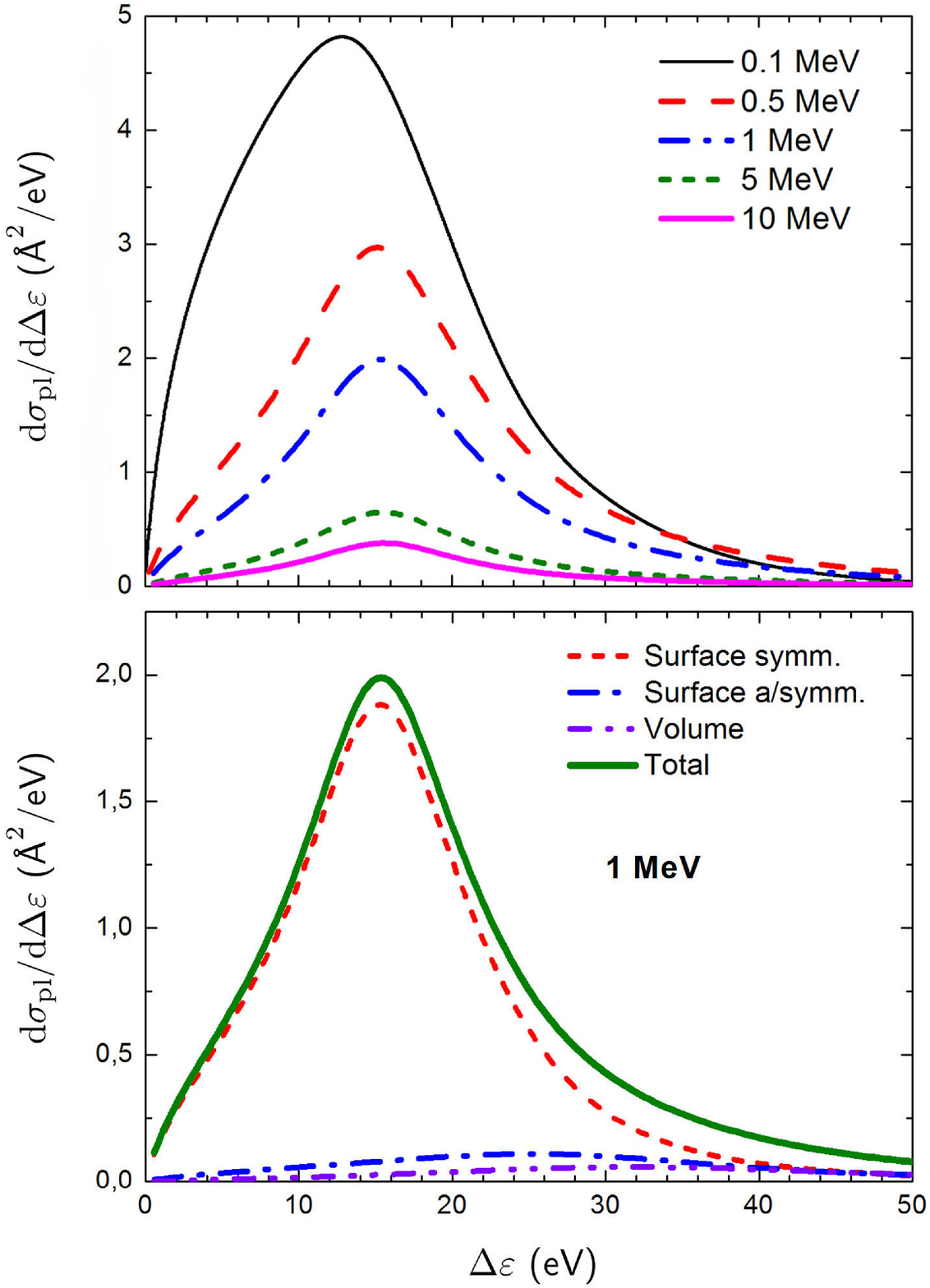}
\caption{
Upper panel: contribution ${\rm d}\sigma_{\rm pl}/{\rm d}\Delta\E$ of the
plasmon excitations to the single differential cross section of C$_{60}$
fullerene irradiated by fast protons of different incident energies as a
function of the energy loss.
Lower panel illustrates the contribution of different plasmon excitations
to the cross section ${\rm d}\sigma_{\rm pl}/{\rm d}\Delta\E$ of C$_{60}$
irradiated by a 1 MeV proton.
}
\label{fig_C60_SDCS}
\end{figure}

\begin{figure*}[htb!]
\centering
\includegraphics[width=0.98\textwidth,clip]{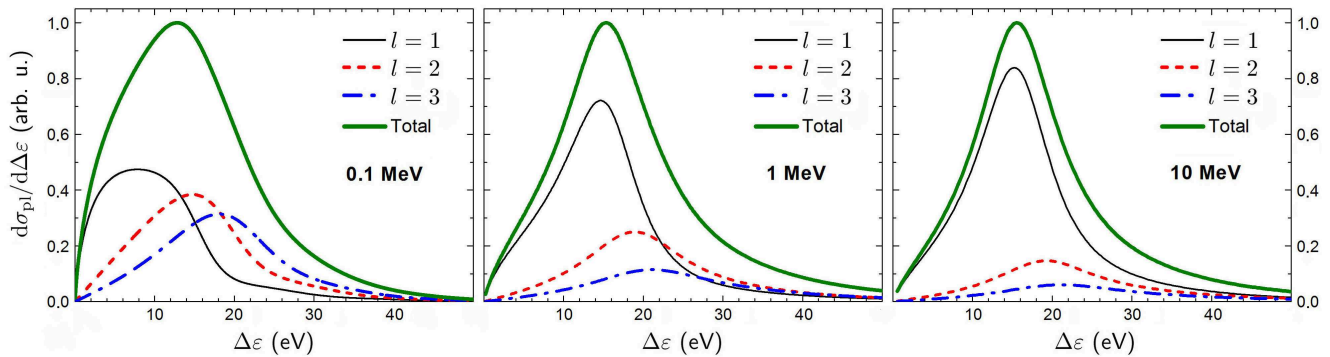}
\caption{Relative contribution of different multipole terms to the single
differential cross section ${\rm d}\sigma_{\rm pl}/{\rm d}\Delta\E$ of C$_{60}$
fullerene irradiated by 0.1, 1, 10~MeV protons as a function of the energy loss.
}
\label{fig_C60_SDCS_2}
\end{figure*}

The upper panel of Figure~\ref{fig_C60_SDCS} illustrates the single differential
cross section ${\rm d}\sigma_{\rm pl}/{\rm d}\Delta\E$ calculated by means of the
PRA for the C$_{60}$ fullerene irradiated by fast protons of different incident energies
as indicated.
The presented spectra comprise contributions of both the surface and volume
plasmon excitations of different angular momenta $l$.
As mentioned in Section~\ref{PRA_theory}, we have accounted for the dipole ($l = 1$),
quadrupole ($l = 2$), and octupole ($l = 3$) plasmon terms because the excitations
with higher angular momentum are formed by single electron transitions rather than
by the collective ones.
The contribution of different plasmon modes to the spectrum of C$_{60}$ irradiated
by a 1~MeV proton is illustrated in the lower panel of Figure~\ref{fig_C60_SDCS}.
The main contribution to the cross section comes from the symmetric mode of the
surface plasmon, whose relative contribution exceeds that of the volume plasmon
by about an order of magnitude.
The similar trend was observed recently studying electron production
by noble metal nanoparticles \cite{Verkhovtsev_2014_NPs, Verkhovtsev_2014_GNP}.
Thus, the leading mechanism of electron production by sensitizing nanoparticles
due to the plasmon excitations should be related to the surface term but not to
the volume one.

Figure~\ref{fig_C60_SDCS} demonstrates that the amplitude and the shape of the
plasmon resonance depend strongly on the kinetic energy of protons.
It was shown previously \cite{Gerchikov_1997_JPhysB.30.4133} that the relative
contributions of the quadrupole and higher multipole terms to the cross
section decrease significantly with an increase of the collision velocity.
At high velocities, the dipole term dominates over the contributions of larger $l$,
since the dipole potential decreases slower at large distances than the higher
multipole potentials.
To illustrate this effect, we have plotted the partial contributions of different
multipole modes which are excited due to irradiation by 0.1, 1, and 10~MeV protons.
These dependencies are presented in Figure~\ref{fig_C60_SDCS_2}.
For the sake of clarity, the cross sections, which represent the sum of three
multipole contributions, have been normalized to unity at the point of maximum.
Thus, one can compare directly the relative contribution of the different terms
to the cross section at different incident energies.
A prominent interplay of the different multipole terms at the lowest incident
energy (left panel) results in a shift in the position of the maximum of the
cross section.

To quantify the production of electrons in collision with a nanoparticle,
we redefine the cross section ${\rm d}\sigma / {\rm d}\Delta\E$ as a
function of the kinetic energy $W$ of emitted electrons.
This quantity is related to the energy loss via $W = \Delta\E - I_p$,
where $I_p$ is the ionization threshold of the system.
The first ionization potential of the C$_{60}$ fullerene approximately
equals to 7.5~eV \cite{Hertel_1992_PhysRevLett.68.784}.

\begin{figure}[ht]
\centering
\includegraphics[width=0.47\textwidth,clip]{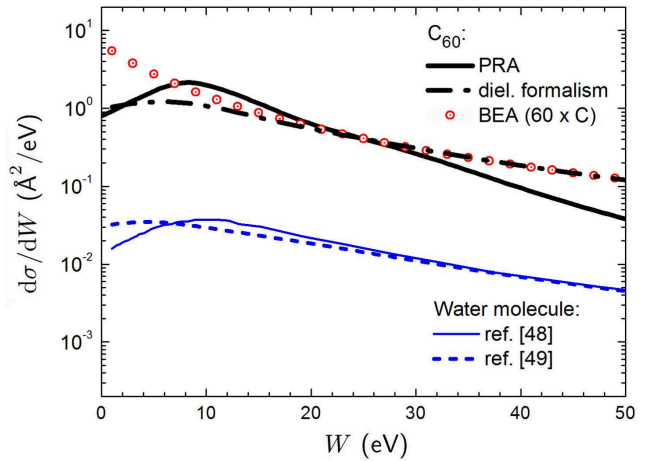}
\caption{Single differential cross section ${\rm d}\sigma/{\rm d}W$ of the
C$_{60}$ fullerene (thick solid and dash-dotted black curves) and of a water
molecule (thin solid and dashed blue curves) irradiated by a 1~MeV proton as
a function of the kinetic energy of emitted electrons.
Thick solid (black) curve illustrates the contribution of the plasmon excitations
to the emission spectrum from C$_{60}$.
Thin solid and dashed (blue) curves represent the results obtained within the
dielectric formalism by Scifoni {\it et al.} \cite{Scifoni_2010_PhysRevE.81.021903}
and de Vera {\it et al.} \cite{deVera_2013_PhysRevLett.110.148104}, respectively.
Symbols represent the cross section of a single C atom calculated by means of
BEA, multiplied by 60.}
\label{fig_C60_SDCS_3}
\end{figure}

Figure \ref{fig_C60_SDCS_3} shows the cross section
${\rm d}\sigma/{\rm d}W$ of C$_{60}$ (thick solid and dash-dotted black curves)
and of a water molecule (thin solid and dashed blue curves) irradiated by a 1~MeV
proton as a function of the kinetic energy of emitted electrons.
The results for water obtained within the dielectric formalism are taken from
references~\cite{Scifoni_2010_PhysRevE.81.021903, deVera_2013_PhysRevLett.110.148104}.
The thick solid curve demonstrates the contribution of the plasmon excitations to
the spectrum of C$_{60}$, ${\rm d}\sigma_{\rm pl}/{\rm d}W$, calculated within the
PRA approach.
The dash-dotted curve represents the results obtained within the dielectric formalism.
In the latter case, we took the experimental optical energy-loss function of
fullerite \cite{Kuzuo_1991_JpnJApplPhys.30.L1817} and calculated the mean binding
energy of the outer-shell electrons.
The binding energies of the valence orbitals of C$_{60}$ were taken from the
{\it ab initio} calculations of Deutsch {\it et al.} \cite{Deutsch_1996_JPhysB.29.5175}.
Symbols show the cross section ${\rm d}\sigma/{\rm d}W$ for the 1~MeV
proton impact of a single carbon atom calculated by means of the binary
encounter approximation (BEA) \cite{Rudd_1992_RevModPhys.64.441, ICRU55_report},
multiplied by 60.
The results of the calculations based on the dielectric formalism agree well
with those within the BEA at the energy of about 20~eV and above.
This indicates that the emission of electrons with kinetic energy of about
several tens of eV takes place via single-electron excitations
of the system.
The plasmon excitations dominate the spectrum at lower energies,
i.e. in the vicinity of the plasmon resonance, while this contribution
drops off at higher energies of emitted electrons.
In the energy range where the plasmons are excited, single-particle effects
give a small contribution as compared to the collective modes.
At higher energies, the collective excitation decays to the incoherent
sum of single-electron excitations.
Note that at lower electron energies (from 1 to approximately 20~eV)
the BEA-based results start to deviate significantly from that of the
dielectric formalism.
This deviation indicates that the BEA is not applicable for the
description of low-energy electron emission, since these electrons are
produced in distant rather than in binary collisions.
In this energy range, the PRA approach better describes the low-energy electron
emission since it accounts for the collective electron effects omitted in
other models.

\subsection{Electron production by a large carbon nanoparticle}

In the previous section, we have calculated the single differential cross section
for an isolated C$_{60}$ molecule within the PRA approach and the dielectric formalism.
Now, we we apply these methods as well as the Monte Carlo scheme to study the
production of secondary electrons by a large solid carbon nanoparticle whose
density corresponds to that of fullerite, the crystalline form of C$_{60}$.

The single differential cross section ${\rm d}\sigma/{\rm d}W$ can be related
to the probability to produce $N$ secondary electrons with kinetic energy $W$,
in the interval $dW$, emitted from a segment $\Delta x$ of the trajectory of a
single ion \cite{Surdutovich_2014_EPJD_Colloquia_Paper, Surdutovich_2007_EPJD.51.63}:
\begin{equation}
\frac{{\rm d}N(W)}{{\rm d}W} = n \Delta x \frac{{\rm d}\sigma}{{\rm d}W} \ ,
\end{equation}

\noindent where $n$ is the atomic density of a system of compounds,
\begin{equation}
n = \frac{\rho}{N_{\rm at} \, m_{\rm at}} \ ,
\end{equation}

\noindent with $\rho$ being the mass density of a target,
$N_{\rm at}$ the number of atoms in the target compound,
and $m_{\rm at}$ the atomic mass.

\begin{figure}[ht]
\centering
\includegraphics[width=0.45\textwidth,clip]{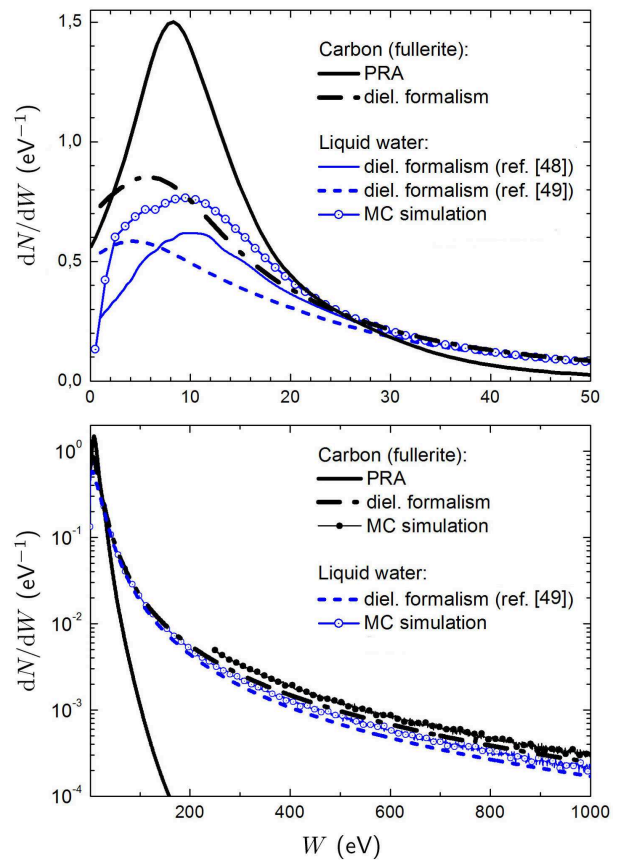}
\caption{Number of electrons per unit energy produced by irradiation
of a 50 nm carbon nanoparticle by a single 1 MeV proton (black curves
and filled circles).
Blue curves represent the number of electron generated in the equivalent
volume of liquid water.
Solid and dashed blue curves represent the results obtained within the
dielectric formalism by Scifoni {\it et al.} \cite{Scifoni_2010_PhysRevE.81.021903}
and de Vera {\it et al.} \cite{deVera_2013_PhysRevLett.110.148104}, respectively.
Open circles illustrate this quantity obtained on the basis of Monte-Carlo
simulations using the Geant4-DNA simulation tool.}
\label{fig_ElProd}
\end{figure}

As a case study, we have considered a nanoparticle of 50~nm in diameter.
In the calculations, we assumed that
(i) C$_{60}$ molecules in fullerite are packed in the fcc crystalline lattice,
and
(ii) a unit cell is composed of four C$_{60}$ molecules.
Knowing the lattice parameter of fullerite, $a = 1.417$~nm, and the mass of
a single carbon atom, $m_{\rm C} = 12$~u, we have calculated the density
of the fullerite crystal:
\begin{equation}
\rho({\rm fullerite}) = \frac{4 \cdot 60\,m_{\rm C}}{a^3} = 1.68 \ {\rm g/cm^3} \ .
\end{equation}

\noindent
Utilizing these values, we have obtained the atomic density of fullerite:
\begin{equation}
n({\rm fullerite}) = \frac{\rho({\rm fullerite})}{ 60 \cdot m_{\rm C}}
= 1.4 \cdot 10^{21}~{\rm cm}^{-3} \ ,
\end{equation}

\noindent
which is by about an order of magnitude smaller than that of water,
$n({\rm water}) = 3.34 \cdot 10^{22}~{\rm cm}^{-3}$.

In Figure~\ref{fig_ElProd}, we compare the electron yield from a 50~nm
spherical carbon nanoparticle and from the equivalent volume of pure water medium.
We have calculated the number of electrons per unit energy produced due to
irradiation by a 1 MeV proton.
Thick black curve represents the contribution of collective electron excitations
estimated by means of the PRA.
The dash-dotted black curve shows the number of electrons estimated by means of
the dielectric formalism.
Filled and open symbols represent the results of the Monte Carlo simulations
carried out by means of the Geant4 tool for the carbon nanoparticle and pure
water medium, respectively.
Thin solid and dashed blue curves represent the results of recent calculation
for liquid water obtained within the dielectric formalism
\cite{Scifoni_2010_PhysRevE.81.021903,deVera_2013_PhysRevLett.110.148104}.
Note that in the Monte Carlo simulations, we did not simulate the crystalline
lattice of fullerite explicitly but the material properties were set by scaling
the density of the Geant4 element carbon according to the calculated density
$\rho({\rm fullerite})$.

Comparative analysis of the spectra at low kinetic energy of emitted electrons
(the upper panel of Figure~\ref{fig_ElProd}) demonstrates that the number of
electrons with the energy of about 10~eV, produced by the carbon nanoparticle
{\it via the plasmon excitation mechanism},
is several times higher than that created in pure water.
The enhancement of the yield of low-energy electrons may increase the probability
of the tumor cell killing due to the double- or multiple strand break of the DNA
\cite{Surdutovich_2014_EPJD_Colloquia_Paper}.
Similar to the case of noble metal nanoparticles
\cite{Herold_2000_IntJRadiatBiol.76.1357, Hainfeld_2004_PhysMedBiol.49.N309,
McMahon_2011_SciRep.1.18, Chen_2006_JNanosciNanotechnol.6.1159,
Porcel_2010_Nanotechnology.21.085103},
the use of carbon-based nanostructures in cancer treatments with ionizing
radiation can thus produce the sensitization effect.
As it was shown recently \cite{Verkhovtsev_2014_NPs, Verkhovtsev_2014_GNP},
the number of electrons with the energy of about a few eV produced by the noble
metal (gold and platinum) nanoparticles via the plasmon excitation mechanism
exceeds that generated in the same volume of liquid water by an order of magnitude.
In the case of a carbon nanoparticle, the electron yield reaches the maximum
at higher electron energies, namely at about 10~eV.
Assuming this, one can consider novel metal-organic sensitizing nanoparticles,
where collective excitations will arise in both parts of the system.
A proper choice of the constituents will allow one to tune the position of the
resonance peaks in the ionization spectra of such systems and, subsequently,
to cover a broader kinetic energy spectrum of electrons emitted from such
nanoparticles.
The fabrication of new, more efficient types of sensitizers would allow one
to significantly advance modern techniques of cancer treatment with ionizing
radiation.

In the case of electrons with higher kinetic energy (the lower panel of
Figure~\ref{fig_ElProd}), the effect done by the carbon nanoparticle
(filled symbols and dash-dotted black curve)
is also more prominent as compared to pure water (open symbols and dashed
blue curve),
as follows from both the calculations based on the dielectric formalism
and the Monte Carlo simulations.
As discussed above, the contribution of the plasmon excitations rapidly
decreases at the energies exceeding approximately 30~eV.
The PRA accounts only for collective electron excitations
that dominate the ionization spectra at low energies.
At higher energies, the plasmons decay into the incoherent sum of
single-electron excitations whose contribution is the most prominent
in this energy region.

\begin{figure}[ht]
\centering
\includegraphics[width=0.45\textwidth,clip]{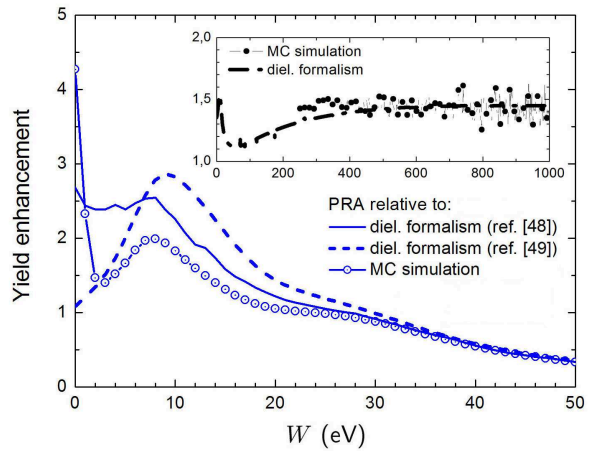}
\caption{
Yield enhancement from the 50~nm carbon nanoparticle as compared to pure water medium.
Solid and dashed blue lines show the enhancement due to the plasmon excitations
as compared to the results obtained within the dielectric formalism by
Scifoni {\it et al.} \cite{Scifoni_2010_PhysRevE.81.021903}
and de Vera {\it et al.} \cite{deVera_2013_PhysRevLett.110.148104}, respectively.
Open symbols illustrate the plasmon-based enhancement compared to the results of
Monte Carlo simulations.
The enhancement estimated solely by means of the dielectric formalism and the
Monte Carlo simulations in a broader kinetic energy range is shown in the inset
by the dash-dotted curve and filled symbols, respectively.}
\label{fig_YieldEnhancement}
\end{figure}

In order to quantify the difference in electron production by the carbon
nanoparticle and by an equivalent volume of pure water, we have calculated
the relative enhancement of the electron yield from the nanoparticle as
compared to water.
This quantity is presented in Figure~\ref{fig_YieldEnhancement}.
The main figure shows the enhancement which was calculated by comparing
the contribution of the plasmon excitations, obtained within the PRA,
to the electron yield from pure water calculated by means of the dielectric
formalism (solid and dashed blue curves) and Monte Carlo simulations (open symbols).
Depending on the data to be chosen as a reference, the collective electron
excitations result in 2 to 3 times greater number of emitted 10~eV electrons
as compared to the case of water.
This effect is less pronounced than the enhancement done by small noble metal
nanoparticles which can produce up to 15-20 times greater number of electrons
via the plasmon decay mechanism as compared to water
\cite{Verkhovtsev_2014_NPs, Verkhovtsev_2014_GNP}.
On the other hand, this enhancement results in an excessive emission of the
very low-energy electrons of about a few eV, while the carbon-based nanoparticle
can enhance the yield of more energetic electrons.
For the sake of completeness, we also demonstrate the enhancement done by the
carbon nanoparticle in a broader kinetic energy range
(see the inset of Figure~\ref{fig_YieldEnhancement}).
For that purpose, we have compared the electron yields from the two systems
calculated by means of the dielectric formalism (dash-dotted curve) and also
from the Monte Carlo simulation (filled symbols).
The two approaches lead to a similar result, namely that the carbon nanoparticle
enhances the number of energetic (of about hundreds of eV up to 1~keV) secondary
electrons by about 50\%.


The analysis performed demonstrates that a single theoretical or numerical
approach does not allow one to properly quantify the secondary electron
yield in a broad kinetic energy range, from a few eV up to a few keV.
Thus, one needs to utilize a combination of different methods to achieve
this goal.
The calculated spectra of secondary electrons can further be used as
the input data for investigation of radiobiological effects by means
of the multiscale approach to the physics of radiation damage with
ions \cite{Surdutovich_2014_EPJD_Colloquia_Paper}.
This approach has the goal of developing knowledge about biodamage at
the nanoscale and molecular level and finding the relation between the
characteristics of incident particles and the resultant biological damage.

\section{Conclusion}

We have analyzed numerically the production of electrons by carbon
nanoparticles irradiated by fast protons.
The study has been carried out by means of the model approaches based
on the plasmon resonance approximation and the dielectric formalism,
as well as by means of Monte Carlo simulations.
It has been demonstrated that due to the prominent collective response
to a time-dependent external electric field, carbon-based nanoparticles
enhance the production of low-energy electrons via the plasmon excitation
mechanism.

The contribution of plasmons to the electron production from a carbon
nanoparticle has been compared to the results of model calculations,
based on the dielectric formalism, as well as to the results of Monte
Carlo simulations for pure water medium.
It has been shown that the number of the low-energy electrons (with the kinetic
energy of about 10~eV) produced by a 50~nm carbon nanoparticle is several times
higher than that emitted from pure water.
Similar to the case of sensitizing metallic nanoparticles, the decay of the
plasmon excitations formed in carbon nanostructures represents an important
mechanism of generation of low-energy electrons.
This observation gives an opportunity to fabricate new types of sensitizers,
composed of the metallic and the organic parts, where the plasmon excitations
will arise in both parts of the system.
As a result, it will become possible to cover a broader kinetic energy range of
electrons emitted from such systems, as compared to currently proposed
nanoagents, and, subsequently, to improve modern techniques of
cancer treatment with ionizing radiation.

\section*{Acknowledgements}

The work was supported by the COST Action MP1002
''Nanoscale insights into Ion Beam Cancer Therapy'' (Nano-IBCT)
and by the FP7 Multi-ITN Project
''Advanced Radiotherapy, Generated by Exploiting Nanoprocesses and Technologies''
(ARGENT) (Grant Agreement n$^{\circ}$608163).


\end{document}